\documentclass[pre,aps,twocolumn,superscriptaddress,showpacs]{revtex4-1}
\usepackage{amssymb}
\usepackage{epsfig}
\usepackage{amsmath}
\usepackage{subfigure}
\usepackage{graphicx}
\usepackage{textcomp}
\usepackage{url}
\usepackage{float}
\usepackage{color}
\usepackage{cases}
\usepackage[normalem]{ulem}

\usepackage{enumitem} 

\usepackage[colorlinks=true, urlcolor=blue, anchorcolor=blue, citecolor=blue,filecolor=blue,linkcolor=blue,menucolor=blue
]{hyperref}

\usepackage{color}
\definecolor{Blue}{rgb}{0.00, 0.00, 0.80}
\definecolor{Red}{rgb}{0.80, 0.00, 0.00}
\definecolor{Green}{rgb}{0.00, 0.50, 0.00}

\usepackage{epsfig}
\usepackage{textcomp}
\usepackage{epstopdf}

\newcommand{\nn}{\nonumber}
\newcommand{\be}{\begin{equation}}
\newcommand{\ee}{\end{equation}}
\newcommand{\bea}{\begin{eqnarray}}
\newcommand{\eea}{\end{eqnarray}}

\begin{document}

\title{Exact short-time height distribution and dynamical phase transition in the relaxation of a Kardar-Parisi-Zhang interface with random initial condition}

\author{Naftali R. Smith}
\email{naftalismith@gmail.com}
\affiliation{Department of Solar Energy and Environmental Physics, Blaustein Institutes for Desert Research, Ben-Gurion University of the Negev,
Sede Boqer Campus, 8499000, Israel}


\begin{abstract}

We consider the relaxation (noise-free) statistics of the one-point height $H=h(x=0,t)$ where $h(x,t)$ is the evolving height of a one-dimensional Kardar-Parisi-Zhang (KPZ) interface, starting from a Brownian (random) initial condition.
We find that, at short times, the distribution of $H$ takes the same scaling form
$-\ln\mathcal{P}\left(H,t\right)=S\left(H\right)/\sqrt{t}$ as the distribution of H for the KPZ interface driven by noise, 
and we find the exact large-deviation function $S(H)$ analytically.
At a critical value $H=H_c$, the second derivative of $S(H)$ jumps, signaling a dynamical phase transition (DPT).
Furthermore, we calculate exactly the most likely history of the interface that leads to a given $H$, and show that the DPT is associated with spontaneous breaking of the mirror symmetry $x \leftrightarrow -x$ of the interface.
In turn, we find that this symmetry breaking is a consequence of the non-convexity of a large-deviation function that is closely related to $S(H)$, and describes a similar problem but in half space. Moreover, the critical point $H_c$ is related to the inflection point of the large-deviation function of the half-space problem.

\end{abstract}

\maketitle

\section{Introduction}

The Kardar-Parisi-Zhang (KPZ) universality class describes fluctuations in a broad range of non-equilibrium systems, including both stochastic interface growth and directed polymers in random media.
In 1+1 dimension, the KPZ equation reads \cite{KPZ}
\begin{equation}
\label{eq:KPZ_dimensional}
\partial_{t}h=\nu\partial_{x}^{2}h+\frac{\lambda}{2}\left(\partial_{x}h\right)^{2}+\sqrt{D}\,\xi(x,t)
\end{equation}
where $h(x,t)$ represents the height of a growing interface as a function of the position $x$ on a substrate and time $t$, and $\xi(x,t)$ is a white (Gaussian) noise with zero mean
and correlation function
\begin{equation}\label{correlator}
\langle\xi(x_{1},t_{1})\xi(x_{2},t_{2})\rangle = \delta(x_{1}-x_{2})\delta(t_{1}-t_{2}).
\end{equation}
At long times, the width of the interface grows as $\sim t^{1/3}$, while the correlation length grows as $\sim t^{2/3}$. These well-known scaling behaviors are hallmarks of the KPZ universality class, and they are shared by many systems and microscopic models that fall within this class \cite{HHZ,Barabasi,Krug,Corwin,QS,S2016,Takeuchi18}.

Height fluctuations in the KPZ equation driven by noise \eqref{correlator} have been studied in great detail. One quantity that has received much attention is the full distribution $P(H,t)$ of the height of the interface at specified point and time $H = h(x=0,t)$. This distribution depends on the initial interface profile $h_0(x) = h(x,t=0)$, and exact representations for $P(H,t)$ were found for three standard initial conditions \citep{SS,CDR,Dotsenko,ACQ, CLD,IS,Borodinetal}: 
These include the flat interface, the ``droplet'' and the stationary initial condition, in which it is assumed that the interface has evolved for a very long time before time $t=0$.
At long times, $t\gg \nu^{5}/(D^{2}\lambda^{4})$, typical fluctuations of $H$ were found to be described by Tracy-Widom \citep{TracyWidom1996} or Baik-Rains \cite{BR} distributions, depending on the initial condition \cite{QS,S2016,Takeuchi18}.

More recently, the short-time ($t \ll \nu^{5}/(D^{2}\lambda^{4})$) height statistics have been studied in some detail. This can be done either by analyzing the short-time asymptotic behavior of the exact representations of $P(H,t)$, or by using the optimal fluctuation method (OFM). The OFM, a generic tool that has been applied in a broad range of systems \cite{Onsager,MSR, Freidlin,Dykman,Graham,
  Falkovich01, GF,EK04, Ikeda2015,Grafke15, bertini2015, Holcman, SmithMeerson2019,
  MeersonSmith2019, AiryDistribution20, BSM1, Grabsch22, MMS22, footnote:OFM}, is based on a saddle-point evaluation of the path integral that corresponds to the stochastic process.
  It was applied to the KPZ equation and closely related systems in Refs. \citep{Mikhailov1991, GurarieMigdal1996,Fogedby1998, Fogedby1999,Nakao2003, KK2007,KK2008,KK2009,Fogedby2009,MKV,KMSparabola,Janas2016,
MeersonSchmidt17, MSV_3d, SMS2018, SKM2018, SmithMeerson2018, SV18, SMV19, ALM19, KLD21, KLD22, SGG22, KLDKMP22}.
For the KPZ equation \eqref{eq:KPZ_dimensional} in 1+1 dimension, it was recently proven rigorously that the OFM becomes asymptotically exact in the short-time limit $t \to 0$ \cite{LinTsai21}.
The short-time behavior of the height statistics, in a proper moving frame \citep{footnote:displacement}, was found to be $-\ln\mathcal{P}\simeq \mathcal{S} \left(H\right)/\sqrt{t}$, where the large-deviation function $\mathcal{S}(H)$ depends on the initial condition. $\mathcal{S}(H)$ was found exactly for the three standard initial conditions \cite{DMRS,LeDoussal2017,SmithMeerson2018}.

Remarkably, it was found in \cite{Janas2016} that the large-deviation function $\mathcal{S} \left(H\right)$ for the stationary interface exhibits a second-order dynamical phase transition (DPT): a jump of the second derivative $\partial_{H}^{2}\mathcal{S}$ at a critical value of $\lambda H=\lambda H_c>0$.
By applying the OFM, it was found that this phase transition is due to a spontaneous breaking of the spatial mirror symmetry $x \leftrightarrow -x$ of the optimal path of the interface conditioned on reaching height $H$. The DPT has been studied in several works \cite{Janas2016, LeDoussal2017, SKM2018, KLD22}. In particular, an effective Landau theory was developed for it in \cite{SKM2018}, in which the Landau free energy was calculated numerically. Moreover, by combining the OFM and the inverse scattering method (which exploits exact integrability of the underlying saddle-point equations), the optimal path was calculated in \cite{KLD22}.
%
Nevertheless, an explicit expression for the optimal path is not so easy to obtain. For instance, in \cite{KLD22}, the optimal path was obtained by using a Fredholm operator inversion formula, which was then evaluated by using the numerical method of Ref.~\cite{KLD21}.
Furthermore, a simple, intuitive understanding of the spontaneous symmetry breaking that causes the DPT is still lacking.

In the KPZ equation, as in many other dynamical systems, the noise drives the system into a steady state.
For such systems, a classical problem is to study the relaxation dynamics after the driving is abruptly turned off. Examples include the study of decay of turbulence \cite{ET00, VV11} or of phase ordering after the temperature is suddenly lowered \cite{Bray94}.
It is thus natural to ask about the relaxation dynamics of a KPZ interface, namely, the statistical behavior of a noiseless ($D=0$) KPZ interface starting from the stationary state.

In fact, in the work \cite{CFS18}, a family of initial conditions $h_0(x)$ for the KPZ equation was introduced, in which $h_0$ is given by a (one-dimensional) Brownian motion as a function of $x$. This family is parametrized by the diffusion coefficient $\mathcal{D}$ of the Brownian motion. For $\mathcal{D} = 0$, one recovers the flat initial condition, and for the particular value $\mathcal{D} = D/4\nu$, one obtains the stationary initial condition.
The relaxation dynamics correspond to the case $D=0$.
In \cite{MeersonSchmidt17} the short-time height statistics were studied for this family of initial conditions. 
The results of \cite{MeersonSchmidt17} were subsequently rigorously proven and extended in \cite{FV21}.
In particular it was shown, analytically and numerically, that the optimal path is of broken symmetry in the tail $\lambda H \to \infty$ of the distribution for any nonzero $\mathcal{D}$, from which it follows that the DPT must be present at any $\mathcal{D} > 0$.
%

The goal of the present work is to study the height statistics of a relaxing KPZ interface, $D=0$. We find  the same short-time scaling $-\ln\mathcal{P}\simeq S \left(H\right)/\sqrt{t}$ as was observed for noisy KPZ dynamics. We calculate the exact rate function $S(H)$ analytically, and also obtain an explicit expression for the optimal history at all times $0 < t < T$.
As one may expect from the results of \cite{MeersonSchmidt17}, we find that our $S(H)$ exhibits a second-order DPT, and calculate the critical height $H_c$ at which it occurs.
Importantly, for our setting, we uncover the reason behind the spontaneous symmetry breaking: as we show, it is a consequence of the non-convexity of a large-deviation function (that we denote below by $s(\mathcal{Q}_R)$) that describes the height statistics in a half-space problem.
Moreover, the critical point of $S(H)$ is related to the inflection point of $s(\mathcal{Q}_R)$.

The remainder of this paper is organized as follows. 
In section \ref{sec:rescaling} we precisely define the problem that we will solve, rescale the variables, and show that the problem decomposes into two decoupled problems, each defined on a half-line.
In section \ref{sec:OFM}, we formulate the OFM problem whose solution gives the short-time large-deviation function $S(H)$. In section \ref{sec:OFMsol} we solve the OFM problem, and show that $S(H)$ exhibits a second-order DPT for which we formulate an effective Landau theory, and show that the DPT is a result of the non-convexity of $s(\mathcal{Q}_R)$.
In section \ref{sec:disc} we summarize and discuss our results, and extend them to several other closely-related settings.
In Appendix \ref{app:asymptotic} we derive the asymptotic behaviors of $S(H)$.

\section{Problem definition and rescaling}

\label{sec:rescaling}

Let us denote by $T$ the time at which we measure the interface height, so $H=h\left(x=0,t=T\right)$.
We rescale space, time and the interface as $x/\sqrt{\nu T}\to x$, $t/T \to t$, $\left|\lambda\right|h/\nu\to h$, so that Eq.~(\ref{eq:KPZ_dimensional}) (with $D=0$) takes the dimensionless form \citep{MKV}
\begin{equation}
\label{eq:KPZ_dimensionless}
\partial_{t}h=\partial_{x}^{2}h-\frac{1}{2}\left(\partial_{x}h\right)^{2},
\end{equation}
where we assume $\lambda<0$ without loss of generality \citep{signlambda}.
The initial interface profile $h(x,t=0)$ is given, in the rescaled variables, by a Brownian motion with diffusion coefficient $\mathcal{D}\lambda^{2}\sqrt{T}/\nu^{3/2}$.

A well-known, useful and remarkable property of the deterministic KPZ equation is that it can be exactly solved through the Hopf-Cole transformation. Indeed, one can check that
\be
Q\left(x,t\right)=e^{-h\left(x,t\right)/2},
\ee
satisfies the diffusion equation $\partial_{t}Q=\partial_{x}^{2}Q$. The latter is then immediately solved,
\be
\label{Qxtsol}
Q\left(x,t\right)=\frac{1}{\sqrt{4\pi t}}\int_{-\infty}^{\infty}Q\left(y,0\right)e^{-\left(x-y\right)^{2}/4t}dy
\ee
yielding an explicit expression for $\mathcal{Q}=Q(0,1)$ (or equivalently, $H$) as a function of the initial condition $h_0(x)$,
\bea
\mathcal{Q}&=&e^{-H/2}=\frac{1}{\sqrt{4\pi}}\int_{-\infty}^{\infty}Q\left(x,0\right)e^{-x^{2}/4}dx\nn\\
&=&\frac{1}{\sqrt{4\pi}}\int_{-\infty}^{\infty}e^{-\left(2h_{0}\left(x\right)+x^{2}\right)/4}dx.
\eea
It is useful to consider the two half spaces, $x<0$ and $x>0$, separately. We write $\mathcal{Q}=\mathcal{Q}_{L}+\mathcal{Q}_{R}$ as a sum of the contributions of each of the two halves of the initial condition,
\bea
&&\mathcal{Q}_{L}=\frac{1}{\sqrt{4\pi}}\int_{-\infty}^{0}e^{-\left(2h_{0}\left(x\right)+x^{2}\right)/4}dx,\\
\label{QRdef}
&&\mathcal{Q}_{R}=\frac{1}{\sqrt{4\pi}}\int_{0}^{\infty}e^{-\left(2h_{0}\left(x\right)+x^{2}\right)/4}dx,
\eea
The two halves of the initial condition, $h_{0}\left(x<0\right)$ and $h_{0}\left(x>0\right)$ are statistically independent. As a result, so are $\mathcal{Q}_{L}$ and $\mathcal{Q}_{R}$.
It therefore suffices to calculate the distributions $p(\cdots)$ of each of $\mathcal{Q}_{L}$ and $\mathcal{Q}_{R}$ (due to mirror symmetry, they are clearly identically distributed) the distribution of $\mathcal{Q}$ being given by their convolution
\be
\label{PQconvolution}
P\left(\mathcal{Q}\right)=\int_{0}^{\infty}d\mathcal{Q}_{R}p\left(\mathcal{Q}_{R}\right)p\left(Q-\mathcal{Q}_{R}\right).
\ee

Let us, therefore, calculate the distribution $p(\mathcal{Q}_R)$. For this, it is sufficient to consider only the right half $x>0$ of the system.
The Hopf-Cole solution to Eq.~\eqref{eq:KPZ_dimensionless} presents us with a major simplification.
The mathematical problem of finding $p\left(\mathcal{Q}_{R}\right)$ can be formulated as that of determining the statistics of the functional \eqref{QRdef} of the random process $h_0(x>0)$ whose statistics are known. Namely, the probability of an initial condition $h_0(x>0)$ is $\propto e^{-s\left[h_{0}\left(x\right)\right]/\epsilon}$, where 
\be
\label{sdef}
s\left[h_{0}\left(x>0\right)\right]=\int_{0}^{\infty}h_{0}'\left(x\right)^{2}dx \, ,
\ee
$\epsilon=4\mathcal{D}\lambda^{2}\sqrt{T}/\nu^{3/2}$,
and the initial condition is pinned at the origin, $h_0(0)=0$.
The (exact) solution to this mathematical problem can formally be written as a path integral over all possible initial conditions $h_0(x>0)$,
\bea
\label{pathIntegral}
&&p\left(\mathcal{Q}_{R}\right)= \nn\\
&&\frac{\int \! \mathbb{D}h_{0}\! \left(x\right)e^{-s\left[h_{0}\left(x\right)\right]/\epsilon}\delta\! \left(\! \mathcal{Q}_{R}\! -\! \frac{1}{\sqrt{4\pi}}\! \int_{0}^{\infty}\! e^{-\left(2h_{0}\left(x\right)+x^{2}\right)/4}dx\! \right)}{\int\mathbb{D}h_{0}\left(x\right)e^{-s\left[h_{0}\left(x\right)\right]/\epsilon}}. \nn\\
\eea

%

\section{Optimal fluctuation method}

\label{sec:OFM}

%

In the short-time limit, which formally corresponds to $\epsilon \to 0$, we use the saddle-point approximation in order to evaluate the path integral \eqref{pathIntegral}. This leads to a minimization problem for the action $s[h_0(x)]$, constrained on a given value of $\mathcal{Q}_R$. We take the latter constraint into account by using a Lagrange multiplier $\Lambda$, i.e., we minimize
\be
\label{sLambdah}
s_{\Lambda}\left[h_{0}\left(x\right)\right]=\int_{0}^{\infty}\left[h_{0}'\left(x\right)^{2}-\frac{\Lambda}{\sqrt{4\pi}}e^{-\left(2h_{0}\left(x\right)+x^{2}\right)/4}\right]dx.
\ee
This functional is to be minimized under the boundary conditions
\be
h_{0}\left(0\right)=0,\quad h_{0}'\left(x\to\infty\right)=0,
\ee
the latter of which is necessary to obtain a finite action $s$.
We emphasize that the variational problem in Eq.~\eqref{sLambdah} doesn't involve time, therefore this version of OFM is much simpler than the field-theory version in which one must minimize a functional of the entire history $h(x,t)$ \cite{MKV}. 
The two ingredients that enable this simplification are the lack of dynamical noise, and the Hopf-Cole solution \eqref{Qxtsol}, the latter enabling one to express $\mathcal{Q}_R$ explicitly as a functional of the initial interface $h_0(x)$.

After solving the OFM minimization problem, one evaluates the action $s$ which is the short-time large-deviation function of $\mathcal{Q}_R$,
\be
\label{pQRscaling}
p\left(\mathcal{Q}_{R},T\right)\sim\exp\left[-\frac{\nu^{3/2}}{4\mathcal{D}\lambda^{2}\sqrt{T}}s\left(\mathcal{Q}_{R}\right)\right]\,.
\ee
The value of $\Lambda$ is ultimately determined by $\mathcal{Q}_R$.
By plugging Eq.~\eqref{pQRscaling} into \eqref{PQconvolution}, one finds that the integral in \eqref{PQconvolution} is dominated by the saddle point, leading to a similar scaling for the original (full-space) problem \cite{footnote:fullspaceOFM}
\be
\label{PQscaling}
P\left(\mathcal{Q},T\right)\sim\exp\left[-\frac{\nu^{3/2}}{4\mathcal{D}\lambda^{2}\sqrt{T}}S\left(\mathcal{Q}=e^{\left|\lambda\right|H/2\nu}\right)\right]\,,
\ee
where
\be
\label{SQmin}
S\left(\mathcal{Q}\right)=\min_{\mathcal{Q}_{R}}\left[s\left(\mathcal{Q}_{R}\right)+s\left(\mathcal{Q}-\mathcal{Q}_{R}\right)\right] \, .
\ee
As we show below, a very interesting feature of this problem is that, despite the mirror symmetry of the KPZ interface, the minimum in \eqref{SQmin} is not always attained at $\mathcal{Q}_{R} = \mathcal{Q} / 2$. This leads to a second-order DPT for $S(\mathcal{Q})$,  accompanied by a spontaneous breaking of the mirror symmetry $x \leftrightarrow -x$ for the optimal path in the full-space problem.

In \cite{MeersonSchmidt17} the more general problem, in which $D > 0$, was studied. Their results are given in terms of a large-deviation function $\mathcal{S}\left(H,\sigma\right)$ where $\sigma=2\sqrt{\nu\mathcal{D}/D}$ is the ratio between the diffusion coefficient $\mathcal{D}$ of the Brownian initial interface and the diffusion coefficient that describes the stationary initial condition.
Our results correspond to the limit $D \to 0$, i.e., the limit $\sigma \to \infty$, and our large-deviation function $S(H)$ is related to that of \cite{MeersonSchmidt17} via
\be
\label{SofHFromMS}
S\left(H\right)=\lim_{\sigma\to\infty}\sigma^{2}\mathcal{S}\left(H,\sigma\right) \, .
\ee
This relation can be derived by comparing our scaling \eqref{PQscaling} with the corresponding one from \cite{MeersonSchmidt17}.
$\mathcal{S}\left(H,\sigma\right)$ was not calculated exactly in \cite{MeersonSchmidt17}, except in certain asymptotic regimes (see below), away from the critical point. In this paper we analytically calculate the exact rate function $S\left(H\right)$ at all $H$, and develop a better understanding of the phase transition.

\section{Solution}

\label{sec:OFMsol}

\subsection{Half-space problem}

\begin{figure}[ht]
\includegraphics[width=0.3\textwidth,clip=]{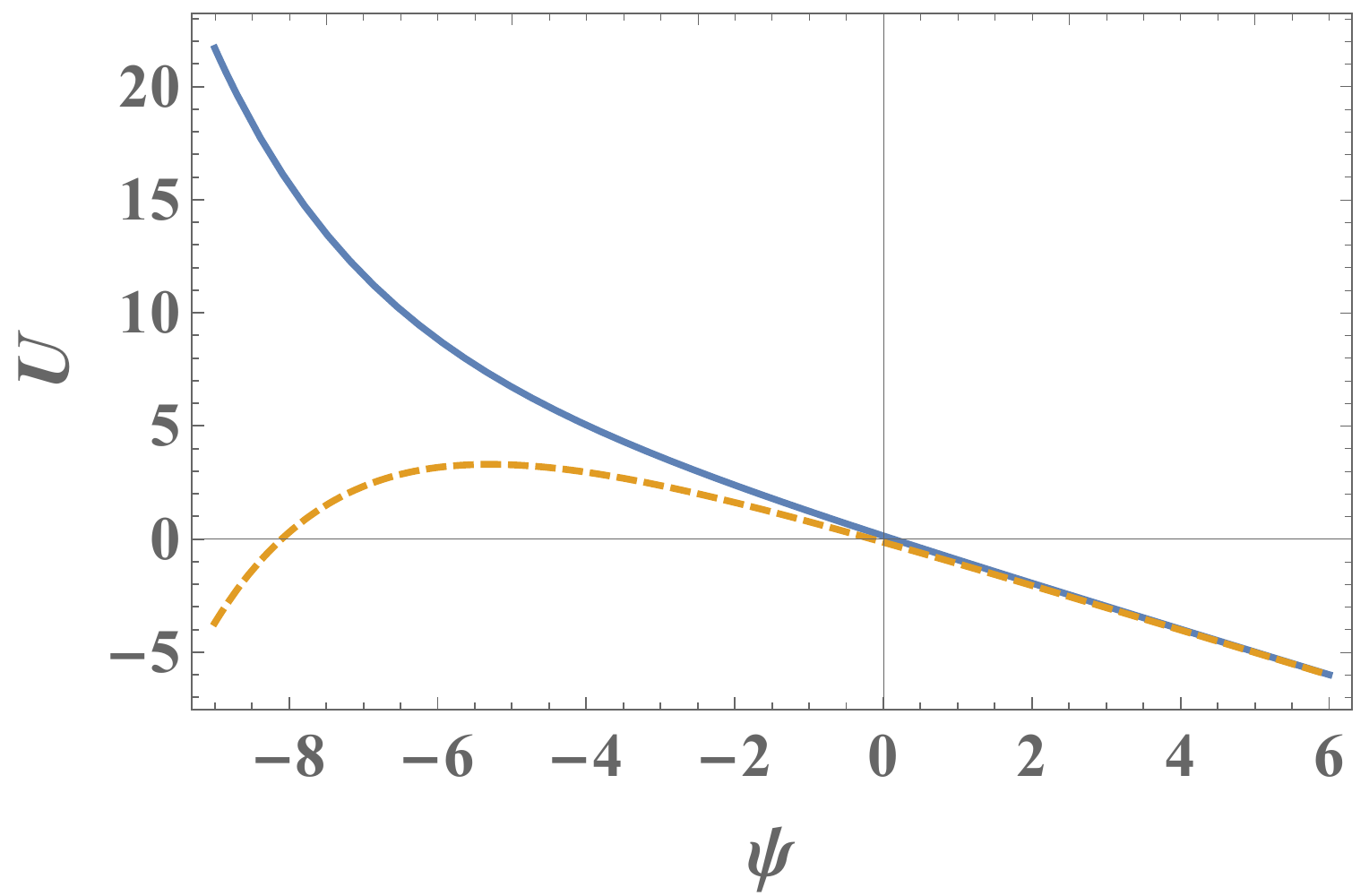}
\caption{The effective potential $U(\psi)$ for $\Lambda = 1$ (solid line) and $\Lambda = -1$ (dashed line).}
\label{fig:Uofpsi}
\end{figure}

We now calculate $s\left(\mathcal{Q}_{R}\right)$ by solving the half-space OFM problem exactly.
It is convenient to rewrite Eq.~\eqref{sLambdah} as a functional of $\psi\left(x\right)=h_{0}\left(x\right)+x^{2}/2$,
\be
s_{\Lambda}=\int_{0}^{\infty}\left[\left(\psi'\left(x\right)-x\right)^{2}-\frac{\Lambda}{\sqrt{4\pi}}e^{-\psi\left(x\right)/2}\right]dx.
\ee
The ensuing Euler-Lagrange equation is
\be
\label{ELpsi}
\psi''\left(x\right)=1+\frac{\Lambda}{4\sqrt{4\pi}}e^{-\psi\left(x\right)/2} \, ,
\ee
and it is to be solved subject to the boundary conditions $\psi(0) = 0$ and
\be
\label{BCxinfty}
\lim_{x\to\infty}h_{0}'\left(x\right)=\lim_{x\to\infty}\left[\psi'\left(x\right)-x\right]=0 \, .
\ee
We solve Eq.~\eqref{ELpsi} by making use of a mechanical analogy: It can be interpreted as Newton's second law of motion, where $\psi$ and $x$ take the roles of position and time respectively, for a particle of unit mass moving in the presence of the potential
\be
\label{Upsidef}
U\left(\psi\right)=-\psi+\frac{\Lambda}{2\sqrt{4\pi}}e^{-\psi/2} \, ,
\ee
see Fig.~\ref{fig:Uofpsi}.
Therefore, the total `energy' of the particle is conserved,
\be
\label{Edef}
\frac{1}{2}\psi'\left(x\right)^{2}-\psi\left(x\right)+\frac{\Lambda}{2\sqrt{4\pi}}e^{-\psi\left(x\right)/2}=E= \text{const}.
\ee
By solving this equation for $\psi'(x)$, we obtain
\be
\label{dpsidx}
\pm\frac{d\psi}{\sqrt{2E+2\psi-\frac{\Lambda}{\sqrt{4\pi}}e^{-\psi/2}}}=dx\, .
\ee

Before we continue, we express $E$ as a function of $\mathcal{Q}_R$ and $\Lambda$.
By evaluating Eq.~\eqref{QRdef} on the minimizer $h_0(x)$ of $s$, we obtain
\bea
\label{QRofpsiprime0}
\mathcal{Q}_{R}&=&\frac{1}{\sqrt{4\pi}}\int_{0}^{\infty}e^{-\psi\left(x\right)/2}dx\nn\\
&=&\frac{4}{\Lambda}\int_{0}^{\infty}\left[\psi''\left(x\right)-1\right]dx\nn\\
&=&\frac{4}{\Lambda}\left[\psi'\left(x\right)-x\right]_{0}^{\infty}=\frac{-4\psi'\left(0\right)}{\Lambda} \, ,
\eea
where, when moving from the first line to the second we used the Euler-Lagrange equation \eqref{ELpsi}, and in the last equality we used the boundary conditions \eqref{BCxinfty} at $x\to\infty$ and $\psi(0)=0$.
On the other hand, $E$ and $\psi'(0)$ can be related by plugging $x=0$ into the `energy' conservation equation \eqref{Edef}, yielding
\be
\label{Eofpsiprime0}
E=\frac{1}{2}\psi'\left(0\right)^{2}+\frac{\Lambda}{2\sqrt{4\pi}} \, ,
\ee
where we again used $\psi(0)=0$.
Eliminating $\psi'(0)$ from Eqs.~\eqref{QRofpsiprime0} and \eqref{Eofpsiprime0}, we obtain
\be
\label{EofQR}
E=\frac{\Lambda^{2}\mathcal{Q}_{R}^{2}}{32}+\frac{\Lambda}{2\sqrt{4\pi}}
\ee

It turns out that $\psi(x)$ behaves a little differently for positive and negative $\Lambda$. We therefore consider these two cases separately, starting from the simpler of the two cases, $\Lambda < 0$.

\subsection{$\Lambda < 0$}

\begin{figure}[ht]
\includegraphics[width=0.49\linewidth,clip=]{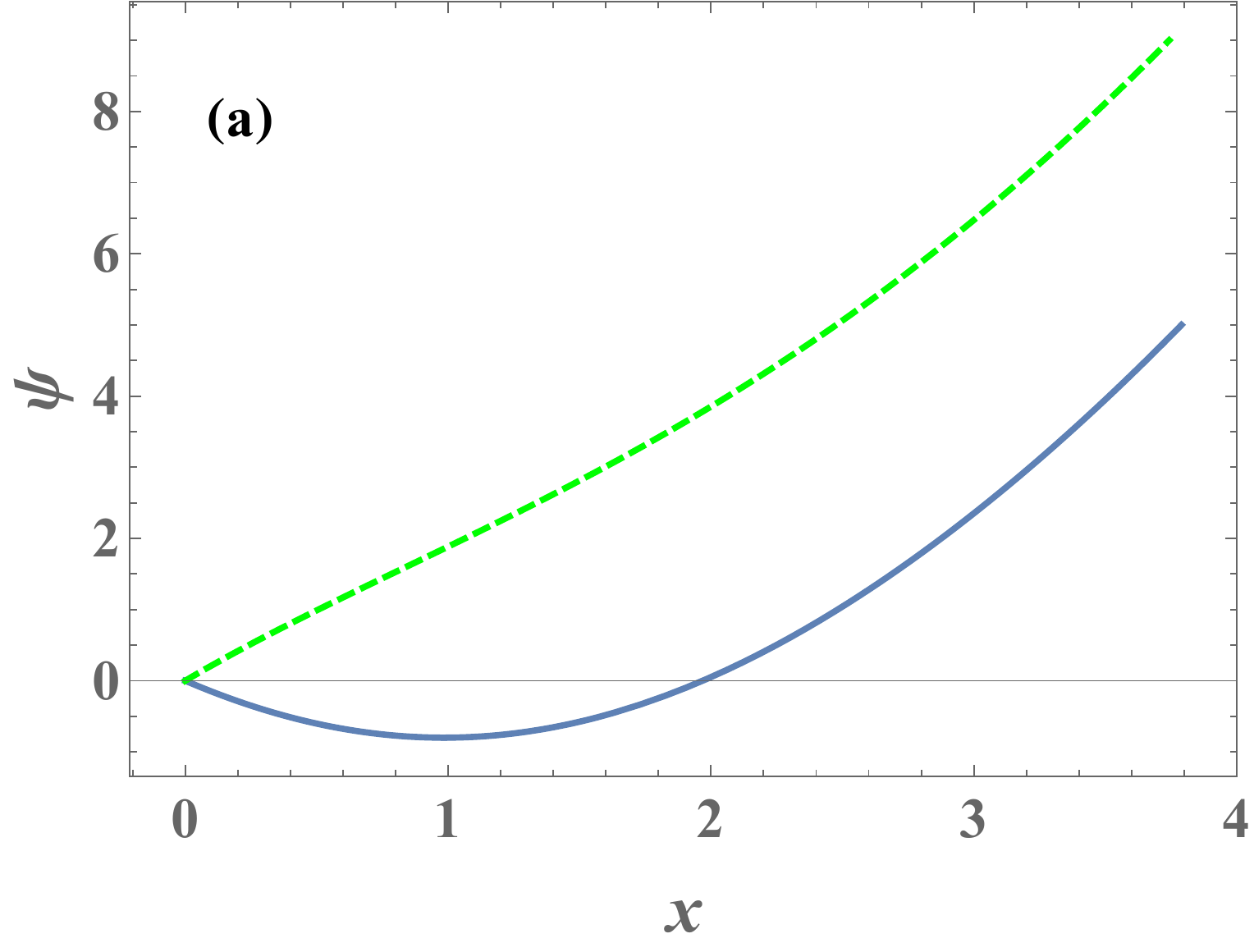}
\includegraphics[width=0.495\linewidth,clip=]{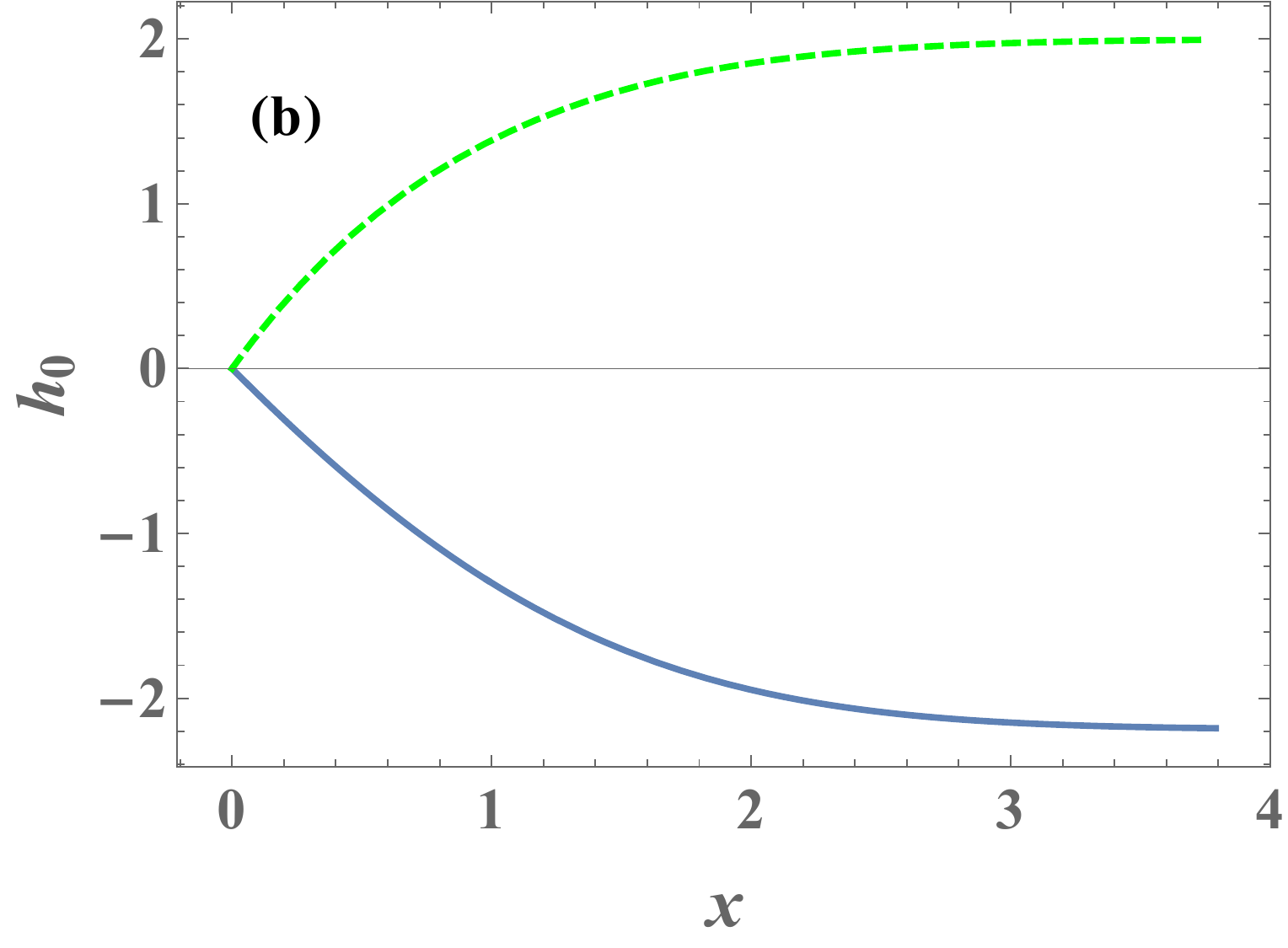}
\caption{$\psi$ vs. $x$ (a) and $h_0(x) = \psi(x) - x^2 /2$ vs. $x$ (b). The dashed lines correspond to the parameters $E=-2$, for which one finds that $\mathcal{Q}_R = 0.28087\dots$ and $\Lambda= -31.264\dots$, and the solid lines correspond to $\psi_{\min} = -0.8$, for which one finds that $\Lambda = 6.6093\dots$, $\mathcal{Q}_R = 0.960154$, and $E= 2.1907\dots$.
As explained in the text, for $\Lambda < 0$, $h_0$ and $\psi$ increase monotonically with $x$, whereas for $\Lambda >0$, $h_0$ is a monotonically decreasing function of $x$ and $\psi(x)$ attains a minimum at  $x=x_{\min} \ne 0$. For the parameters corresponding to the solid lines, $x_{\min} = 0.98412\dots$.}
\label{fig:psiOfx}
\end{figure}

The case $\Lambda<0$ corresponds to $\mathcal{Q}_{R}<1/2$. This case is relatively simple, because it turns out that $\psi'(x)$ is positive at all $x$, as is $h_0'(x)$.
Taking the plus sign in Eq.~\eqref{dpsidx}, we immediately integrate the equation with the boundary condition $\psi\left(0\right)=0$ to get
\be
\label{xOfPsiNegativeLambda}
\int_{0}^{\psi}\frac{d\phi}{\sqrt{2E+2\phi-\frac{\Lambda}{\sqrt{4\pi}}e^{-\phi/2}}}=x \, ,
\ee
see Fig.~\ref{fig:psiOfx} (a).

\begin{figure}[ht]
\includegraphics[width=0.3\textwidth,clip=]{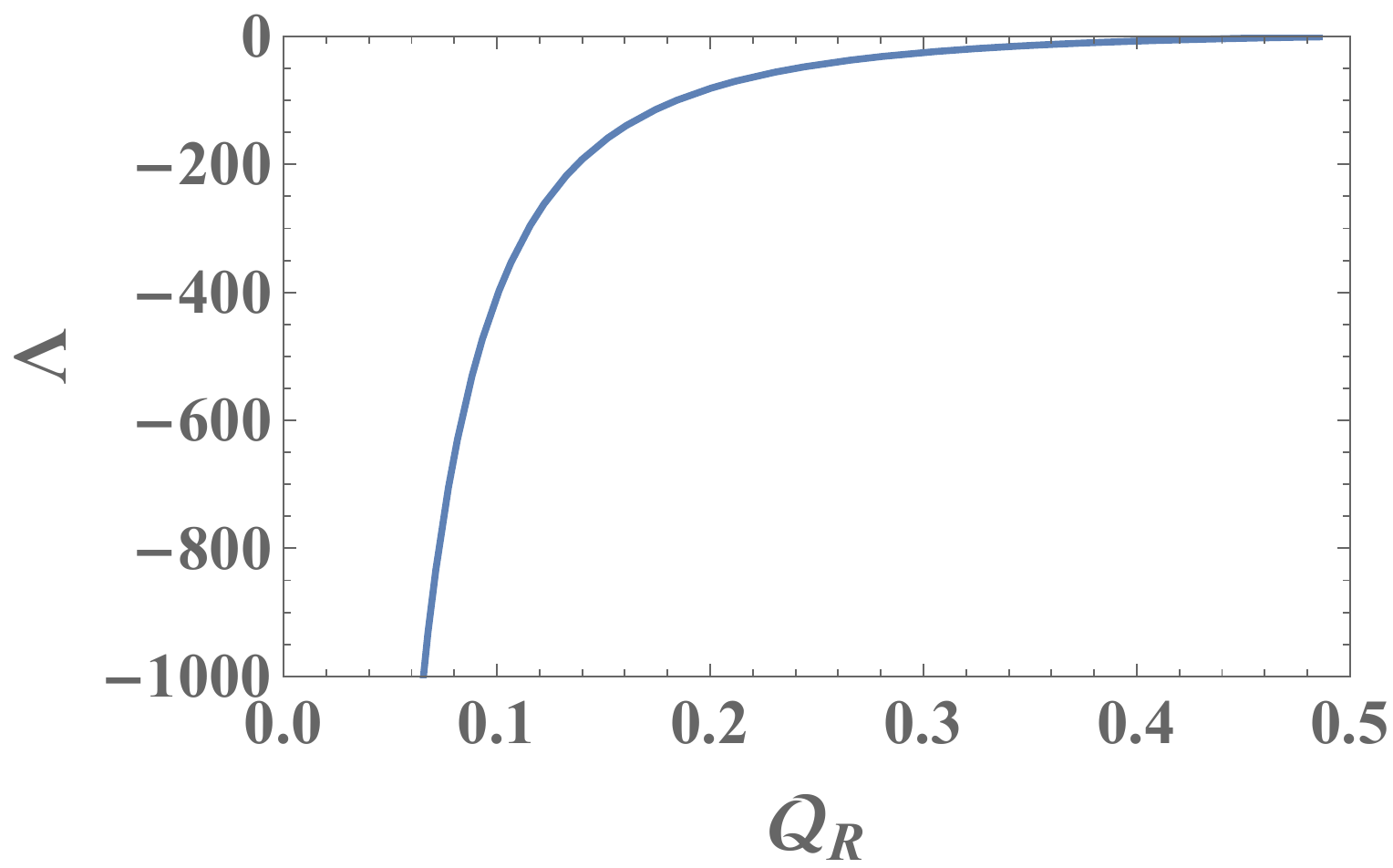}
\caption{$\Lambda$ vs. $\mathcal{Q}_R$ for $\mathcal{Q}_R < 1/2$, which is obtained by numerically solving Eq.~\eqref{QRofLambda}.}
\label{fig:LambdaNeg}
\end{figure}

We are now able to determine the `energy' $E$ as a function of $\Lambda$ [which, through Eq.~\eqref{EofQR}, then gives us $\mathcal{Q}_R$ as a function of $\Lambda$]. This is done by requiring the boundary condition \eqref{BCxinfty}.
By analyzing Eq.~\eqref{xOfPsiNegativeLambda}, we see that its leading-order behavior at $\psi \to \infty$ is, in general,
\be
\label{pisCx1}
\sqrt{2\psi}+C\simeq x \, ,
\ee
where $C$ depends on $E$ and $\Lambda$, and is given by
\be
C=\int_{0}^{\infty}\left(\frac{1}{\sqrt{2E+2\phi-\frac{\Lambda}{\sqrt{4\pi}}e^{-\phi/2}}}-\frac{1}{\sqrt{2\phi}}\right)d\phi \, .
\ee
However, Eq.~\eqref{pisCx1} yields the asymptotic behavior
\be
\psi\left(x\to\infty\right)\simeq\frac{x^{2}}{2}-Cx \, ,
\ee
so
\be
\lim_{x\to\infty}\left[\psi'\left(x\right)-x\right]=-C \, .
\ee
Thus, using the boundary condition \eqref{BCxinfty}, we find that $E$ is determined as a function of $\Lambda$ through the requirement $C=0$, which in turn, using the relation \eqref{EofQR}, yields
\be
\label{QRofLambda}
\int_{0}^{\infty}\!\!\left[\frac{1}{\sqrt{\frac{\Lambda^{2}\mathcal{Q}_{R}^{2}}{16}+2\phi+\frac{\Lambda}{\sqrt{4\pi}}\left(\frac{1}{32}-e^{-\phi/2}\right)}}-\frac{1}{\sqrt{2\phi}}\right] \! d\phi=0.
\ee
Eq.~\eqref{QRofLambda} can be solved numerically to yield $\Lambda$ as a function of $\mathcal{Q}_R$ for $\Lambda < 0$, see Fig.~\ref{fig:LambdaNeg}.

\subsection{$\Lambda > 0$}

The case $\Lambda > 0$ is a little more involved from a technical point of view, because it turns out that $\psi'(0) < 0$, so that $\psi(x)$ is not monotonic: It attains its global minimum $\psi_{\min}$ at a nonzero point $x_{\min} > 0$, see Fig.~\eqref{fig:psiOfx} (a).
As a result, the solution given in the previous subsection does not hold for $\Lambda > 0$.
In Eq.~\eqref{dpsidx} we must take the minus sign for $0 < x < x_{\min}$, and the plus sign for $x > x_{\min}$ which, after integration, yields
\be
\label{xOfPsiPositiveLambda}
x=\begin{cases}
\int_{\psi}^{0}\frac{d\phi}{\sqrt{2E+2\phi-\frac{\Lambda}{\sqrt{4\pi}}e^{-\phi/2}}}, & 0<x<x_{\min},\\[3mm]
x_{\min}+\int_{\psi_{\min}}^{\psi}\frac{d\phi}{\sqrt{2E+2\phi-\frac{\Lambda}{\sqrt{4\pi}}e^{-\phi/2}}}, & x>x_{\min},
\end{cases}
\ee
where
\be
\label{xmindef}
x_{\min}=\int_{\psi_{\min}}^{0}\frac{d\phi}{\sqrt{2E+2\phi-\frac{\Lambda}{\sqrt{4\pi}}e^{-\phi/2}}} \, .
\ee
Before continuing, we can relate $\psi_{\min}$ to $E$ by using the `energy' conservation \eqref{Edef} at $x=x_{\min}$, together with $\psi'(x_{\min}) = 0$. This yields
\be
\label{Eofpsimin}
E=-\psi_{\min}+\frac{\Lambda}{2\sqrt{4\pi}}e^{-\psi_{\min}/2} \, .
\ee

As we did in the case $\Lambda < 0$, we now must use the boundary condition in order to determine $\Lambda$ as a function of $\mathcal{Q}_R$.
Our solution \eqref{xOfPsiPositiveLambda} again yields the asymptotic behavior $\sqrt{2\psi}+C\simeq x \, ,$ in the large-$x$ (or equivalently, large-$\psi$) limit, however, $C$ is now given by
\be
C=x_{\min}+\lim_{\psi\to\infty} \! \left(\int_{\psi_{\min}}^{\psi}  \!  \frac{d\phi}{\sqrt{2E+2\phi-\frac{\Lambda}{\sqrt{4\pi}}e^{-\phi/2}}}-\sqrt{2\psi}\right)  \! .
\ee
We now must require $C=0$ (as in the analysis for the case $\Lambda < 0$). This requirement yields the equation
\bea
\label{LambdaOfpsimin}
&& 2\int_{\psi_{\min}}^{0}\frac{d\phi}{\sqrt{2E\left(\psi_{\min}\right)+2\phi-\frac{\Lambda}{\sqrt{4\pi}}e^{-\phi/2}}} \nn\\
&&+\int_{0}^{\infty}\left(\frac{1}{\sqrt{2E\left(\psi_{\min}\right)+2\phi-\frac{\Lambda}{\sqrt{4\pi}}e^{-\phi/2}}}-\frac{1}{\sqrt{2\phi}}\right)d\phi=0 \nn\\
\eea
where we used Eq.~\eqref{xmindef}.
Finally, after plugging in $E\left(\psi_{\min}\right)$ from Eq.~\eqref{Eofpsimin}, Eq.~\eqref{LambdaOfpsimin} relates $\psi_{\min}$ and $\Lambda$ and can be solved numerically to obtain one of them as a function of the other.

\begin{figure}[ht]
\includegraphics[width=0.3\textwidth,clip=]{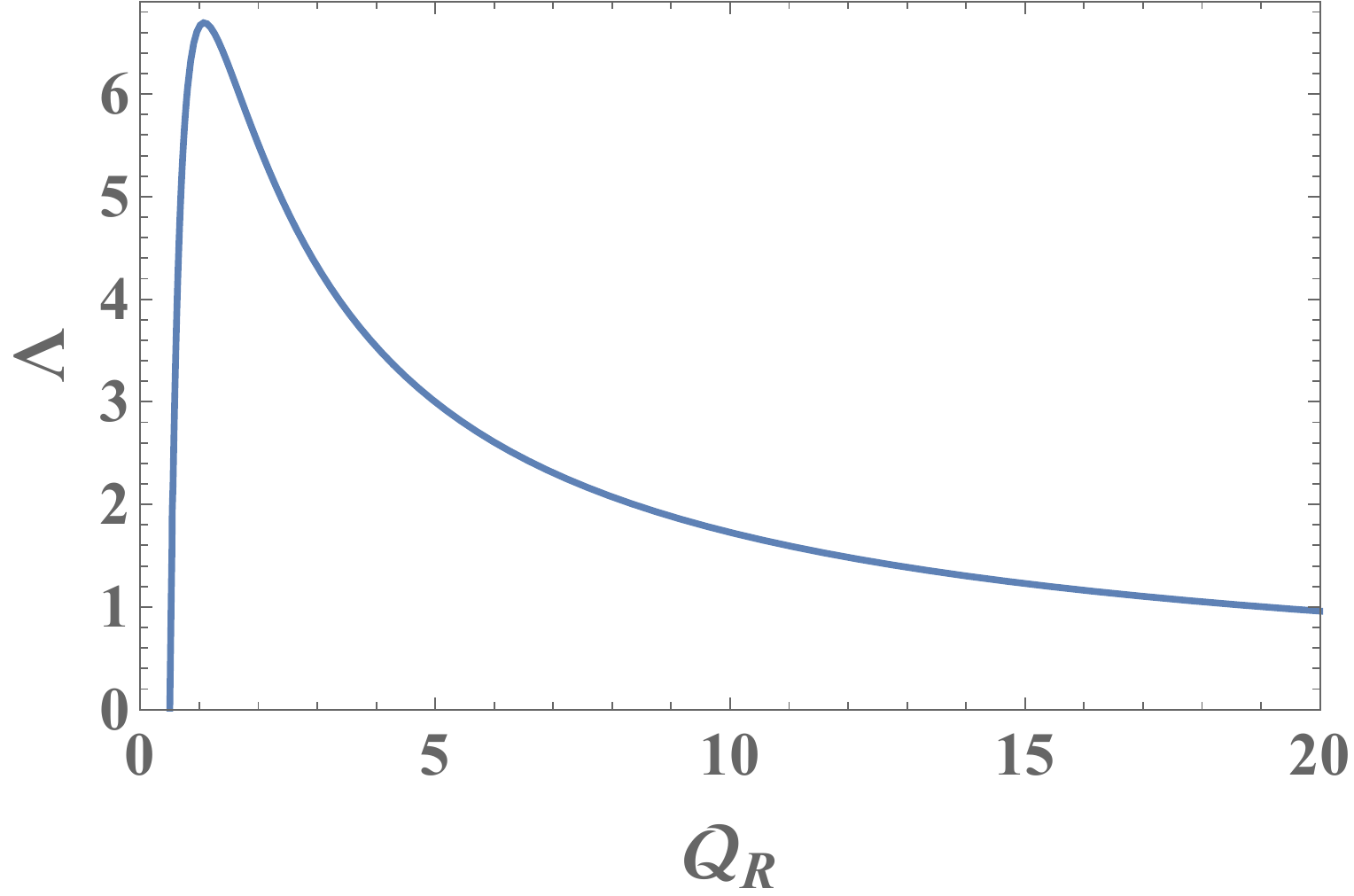}
\caption{$\Lambda$ vs. $\mathcal{Q}_R$ for $\mathcal{Q}_R > 1/2$. This plot is obtained parametrically as $\Lambda = \Lambda(\psi_{\min})$ and $\mathcal{Q}_R = \mathcal{Q}_R(\psi_{\min})$ from Eqs.~\eqref{LambdaOfpsimin} and \eqref{QROfpsimin} respectively, as explained in the text. Importantly, the function $\Lambda(\mathcal{Q}_R)$ is not monotonic. It reaches its maximum at $\mathcal{Q}_{R} = \mathcal{Q}_{R,c} = 1.082\dots$.}
\label{fig:LambdaPos}
\end{figure}

$\Lambda$ is now obtained as a function of $\mathcal{Q}_R$ in a parametric form, $\Lambda = \Lambda(\psi_{\min})$ and $\mathcal{Q}_R = \mathcal{Q}_R(\psi_{\min})$, as follows. Given $\psi_{\min}$, one numerically solves Eq.~\eqref{LambdaOfpsimin} for $\Lambda$, and in addition, calculates $\mathcal{Q}_R$ by using Eqs.~\eqref{Eofpsimin} and \eqref{EofQR}, which together yield
\bea
\label{QROfpsimin}
&&\mathcal{Q}_{R}=\sqrt{\frac{32E}{\Lambda^{2}}-\frac{1}{2\sqrt{4\pi}\,\Lambda}}\nn\\
&&=\sqrt{\frac{32}{\Lambda^{2}}\left(-\psi_{\min}+\frac{\Lambda}{2\sqrt{4\pi}}e^{-\psi_{\min}/2}\right)-\frac{1}{2\sqrt{4\pi}\,\Lambda}} \, .
\eea
$\Lambda(\mathcal{Q}_R)$ is plotted in Fig.~\ref{fig:LambdaPos}. Interestingly, it is not a monotonically increasing function. It grows at small $\mathcal{Q}_R$ until it reaches a (global) maximum at $\mathcal{Q}_{R} = \mathcal{Q}_{R,c} = 1.082\dots$, and then decays until eventually $\Lambda\left(\mathcal{Q}_{R}\to\infty\right)\to0$.

\subsection{Large-deviation function $s(\mathcal{Q}_R)$ and optimal initial condition $h_0(x)$ for the half-space problem}

\begin{figure}[ht]
\includegraphics[width=0.6\linewidth,clip=]{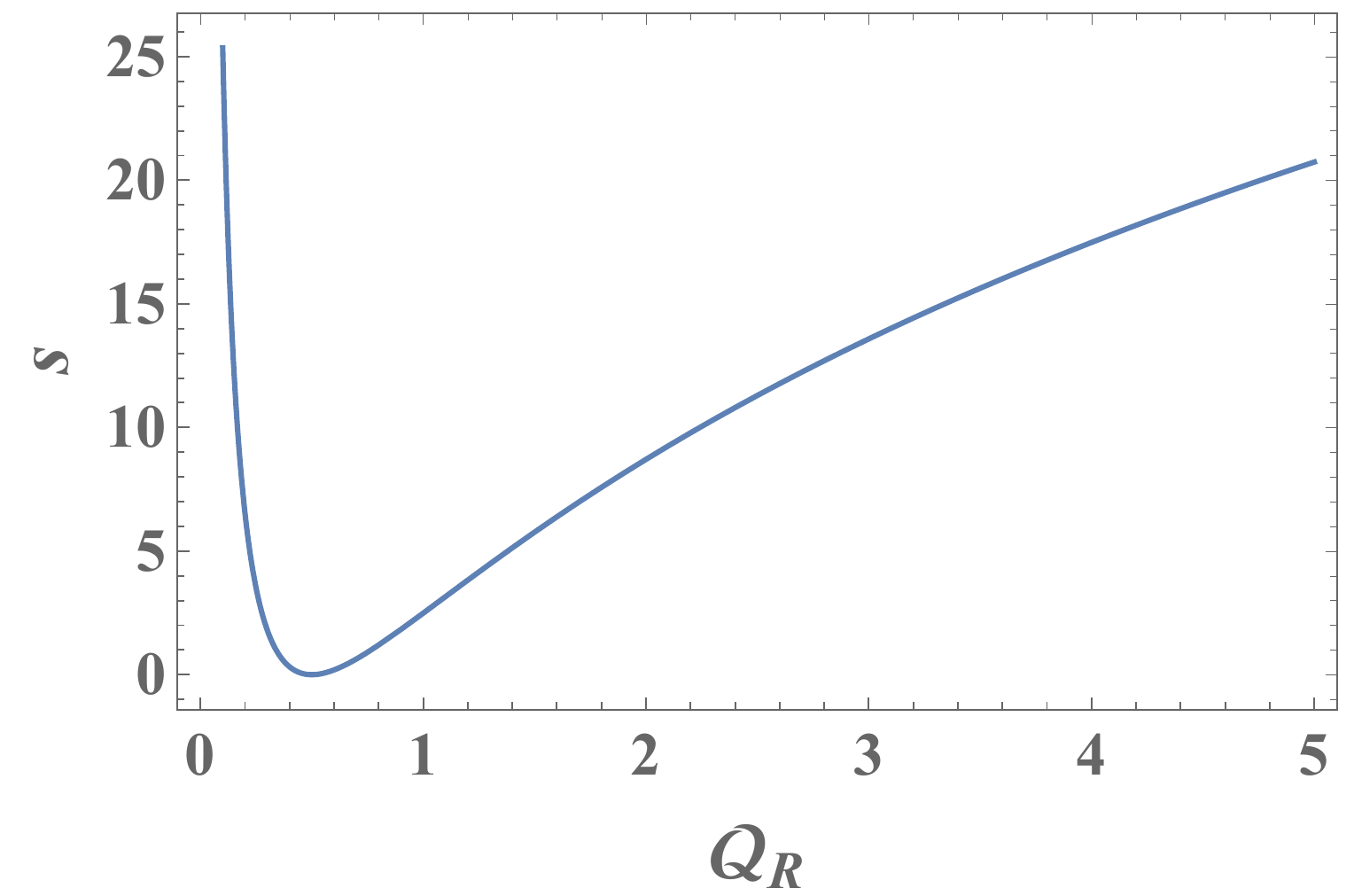}
\caption{The large-deviation function $s(\mathcal{Q}_R)$. Importantly, it is not convex: It has an inflection point at $\mathcal{Q}_{R} = \mathcal{Q}_{R,c} = 1.082\dots$.}
\label{fig:sofQR}
\end{figure}

Let us now calculate the large-deviation function $s(\mathcal{Q}_R)$ at $\Lambda < 0$.
One way of calculating $s$ is to evaluate the integral \eqref{sdef} on the minimizer $h_0(x)$. However, there is a useful shortcut, which makes use of the relation
\be
\frac{ds_{R}}{d\mathcal{Q}_{R}}=\Lambda \, .
\ee
This relation follows from the fact that $\Lambda$ and $\mathcal{Q}_R$ are conjugate variables, see \textit{e.g.} Ref. \cite{Vivoetal}.
Therefore, once we have obtained the function $\Lambda(\mathcal{Q}_R)$, its integral with respect to $\mathcal{Q}_R$ gives us the rate function $s(\mathcal{Q}_R)$ that we are after, see Fig.~\ref{fig:sofQR}.

From the relation $ds/d\mathcal{Q}_{R} = \Lambda$, it also follows that $s$ is a nonconvex function of $\mathcal{Q}_{R}$, with an inflection point at $\mathcal{Q}_{R} = \mathcal{Q}_{R,c}$. This is a relatively atypical situation in large-deviation theory, and as we show below, it has important consequences regarding the full-space problem, and it ultimately gives rise to the phase transition of $S(H)$.

Finally, one can evaluate the optimal initial condition $h_0(x)$ conditioned on a given $\mathcal{Q}_{R}$ exactly. For a given $\Lambda < 0$, one first calculates $\mathcal{Q}_R$ as described above and then uses Eq.~\eqref{EofQR} to calculate $E$. Eq.~\eqref{xOfPsiNegativeLambda} then yields $\psi(x)$.
This yields $h_0(x)$, through $h_0(x) = \psi(x) - x^2 /2$.
For $\Lambda > 0$ the procedure is modified as follows: In order to obtain $E$, one solves Eq.~\eqref{LambdaOfpsimin} for $\psi_{\min}$, and then computes $E\left(\psi_{\min}\right)$ from Eq.~\eqref{Eofpsimin}. $\psi(x)$ is then obtained from Eq.~\eqref{xOfPsiPositiveLambda}.
Optimal initial conditions $h_0(x>0)$ are plotted in Fig.~\ref{fig:psiOfx} for $\Lambda = \pm1$.

\subsection{Full-space problem}

\begin{figure}[ht]
\includegraphics[width=0.485\linewidth,clip=]{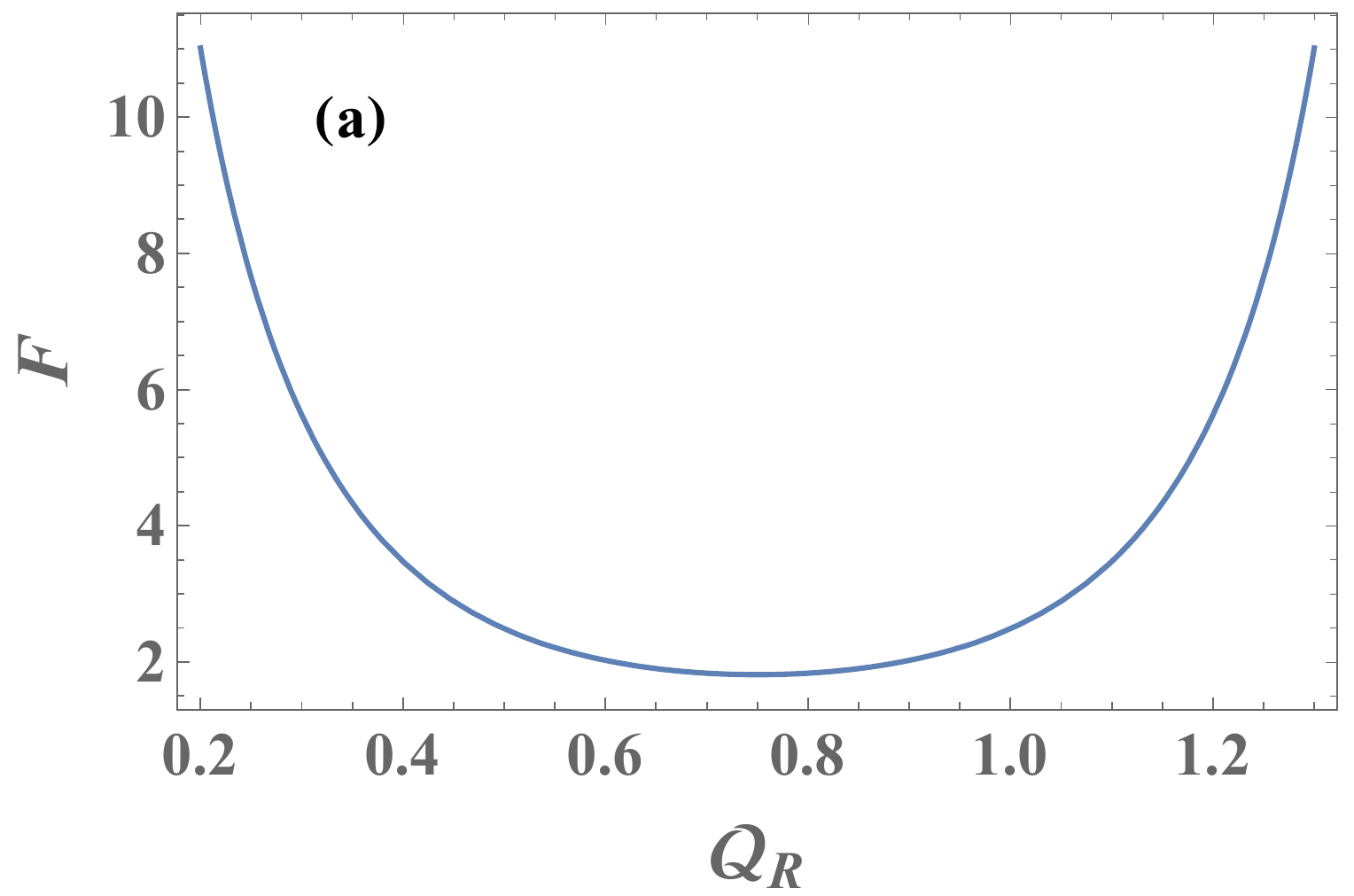}
\includegraphics[width=0.5\linewidth,clip=]{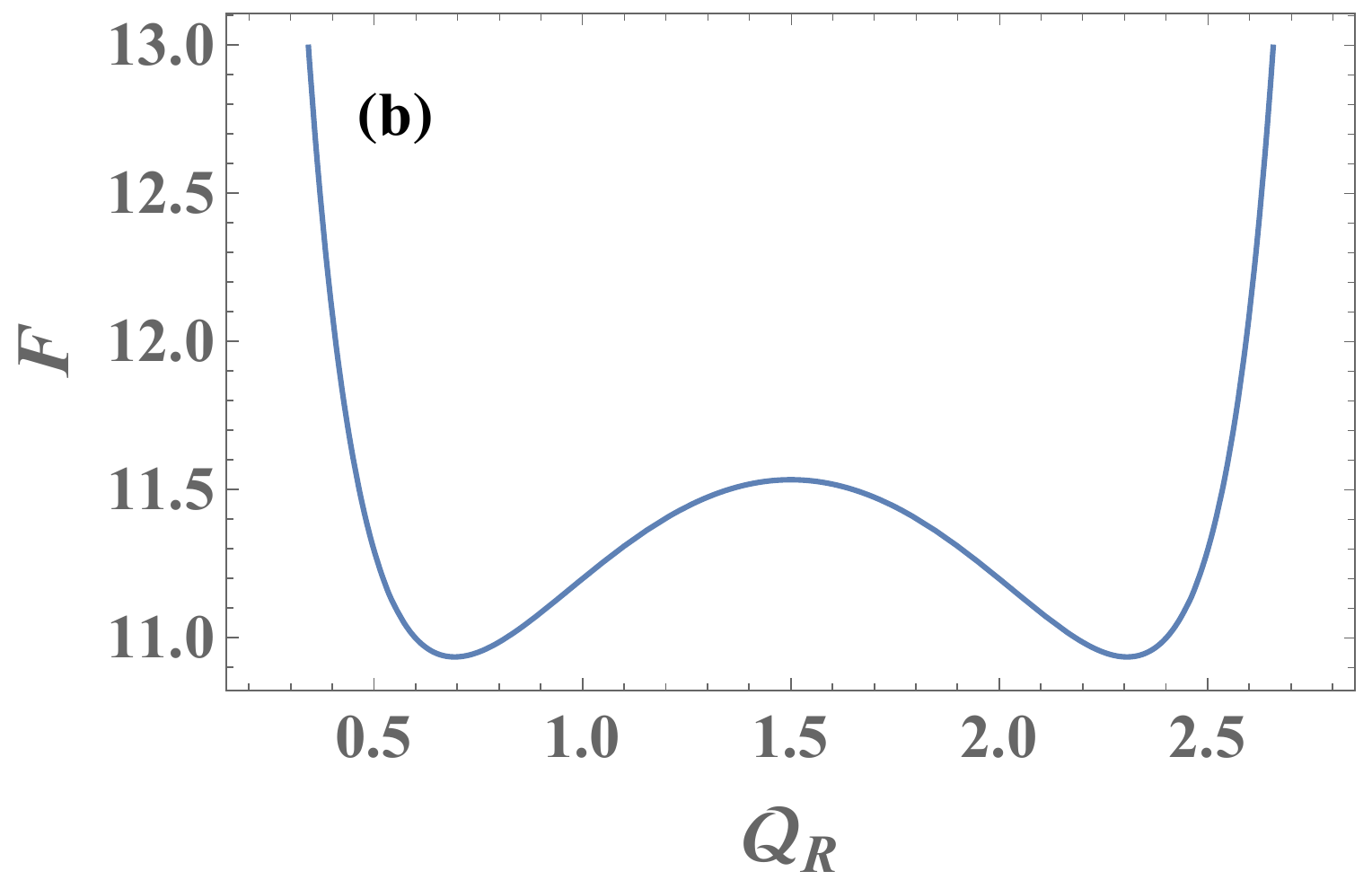}
\caption{$F$ vs. $\mathcal{Q}_R$ for $\mathcal{Q} = 3/2 < \mathcal{Q}_{c}$ (a) and $\mathcal{Q} = 3 > \mathcal{Q}_{c}$ (b). At $\mathcal{Q} < \mathcal{Q}_{c}$, $F$ has a single minimum at $\mathcal{Q}_R = \mathcal{Q}/2$. At $\mathcal{Q} > \mathcal{Q}_{c}$, the point $\mathcal{Q}_R = \mathcal{Q}/2$ becomes a local maximum of $F$, and instead there are two nontrivial minima that correspond to symmetry-broken optimal histories $h(x,t)$ of the interface.}
\label{fig:FofQR}
\end{figure}

Having obtained the exact large-deviation function $s(\mathcal{Q}_R)$, it is now straightforward to calculate $S(\mathcal{Q})$ by using Eq.~\eqref{SQmin}.
It turns out that there is a DPT at a critical value $\mathcal{Q}_{c} = 2\mathcal{Q}_{R,c} = 2.165\dots$, as we now explain.
The minimizer in \eqref{SQmin} is found by requiring the derivative of the function
\be
\label{sQrExpr}
F\left(\mathcal{Q},\mathcal{Q}_{R}\right) = s\left(\mathcal{Q}_{R}\right)+s\left(\mathcal{Q}-\mathcal{Q}_{R}\right)
\ee
 with respect to $\mathcal{Q}_R$ to vanish. This requirement leads to the equation
\be
\label{LambdaQRLambdaQL}
\Lambda\left(\mathcal{Q}_{R}\right)=\Lambda\left(\mathcal{Q}-\mathcal{Q}_{R}\right),
\ee
where we used the relation $ds/d\mathcal{Q}_{R}=\Lambda$.
One obvious solution to this equation is $\mathcal{Q}_R = \mathcal{Q} / 2$. At $\mathcal{Q} < \mathcal{Q}_{c}$, $\mathcal{Q}_R = \mathcal{Q} / 2$ is indeed the minimizer in Eq.~\eqref{SQmin}, see Fig.~\ref{fig:FofQR} (a). Indeed, one finds that the second derivative of $F$ with respect to $\mathcal{Q}_R$ at $\mathcal{Q}_R =\mathcal{Q}/2$ is 
\be
\label{Ftagtag}
\left.\frac{\partial^{2}F}{\partial\mathcal{Q}_{R}^{2}}\right|_{\mathcal{Q}_{R}=\mathcal{Q}/2}=2\Lambda'\left(\frac{\mathcal{Q}}{2}\right) \, ,
\ee
which, for $\mathcal{Q} < \mathcal{Q}_{c}$, is positive.
As a result, $\mathcal{Q}_{L}=\mathcal{Q}-\mathcal{Q}_{R}=\mathcal{Q}/2$, and from Eq.~\eqref{SQmin} we obtain
\be
\label{SofQtwices}
S\left(\mathcal{Q}\right)=2s\left(\mathcal{Q}_{R}=\frac{\mathcal{Q}}{2}\right) \, .
\ee
This the symmetric phase, in which the optimal interface (in the full-space problem) has the mirror symmetry $h(x,t) = h(-x,t)$.

Remarkably, at $\mathcal{Q} > \mathcal{Q}_{c}$, the situation is very different.
The second derivative \eqref{Ftagtag} becomes negative, so
$\mathcal{Q}_R = \mathcal{Q} / 2$ becomes a local \emph{maximum} of $F$ (as a function of $\mathcal{Q}_R$), see Fig.~\ref{fig:FofQR} (b).
Eq.~\eqref{LambdaQRLambdaQL} has two additional solutions, which turn out to be the minimizers in
Eq.~\eqref{SQmin}. They represent two symmetry-broken solutions for the optimal interface that are mirror-images of each other, with the same action.
Mathematically, the existence of solutions to Eq.~\eqref{LambdaQRLambdaQL} with $\mathcal{Q}_{R} \ne \mathcal{Q} - \mathcal{Q}_R$ becomes possible because $\Lambda$ is not a one-to-one function of $\mathcal{Q}_R$, see Fig.~\ref{fig:LambdaPos}. This property  is related to the non-convexity of $s(\mathcal{Q}_R)$, due to the relation $ds/d\mathcal{Q}_{R}=\Lambda$.
So, to summarize, in the symmetry broken phase $\mathcal{Q} > \mathcal{Q}_{c}$, $S(\mathcal{Q})$ is given by
\be
S\left(\mathcal{Q}\right)=s\left(\mathcal{Q}_{R}\right)+s\left(\mathcal{Q}-\mathcal{Q}_{R}\right)
\ee
where $\mathcal{Q}_{R} \ne \mathcal{Q}/2$ satisfies $\Lambda\left(\mathcal{Q}_{R}\right)=\Lambda\left(\mathcal{Q}-\mathcal{Q}_{R}\right)$.

\begin{figure}[ht]
\includegraphics[width=0.492\linewidth,clip=]{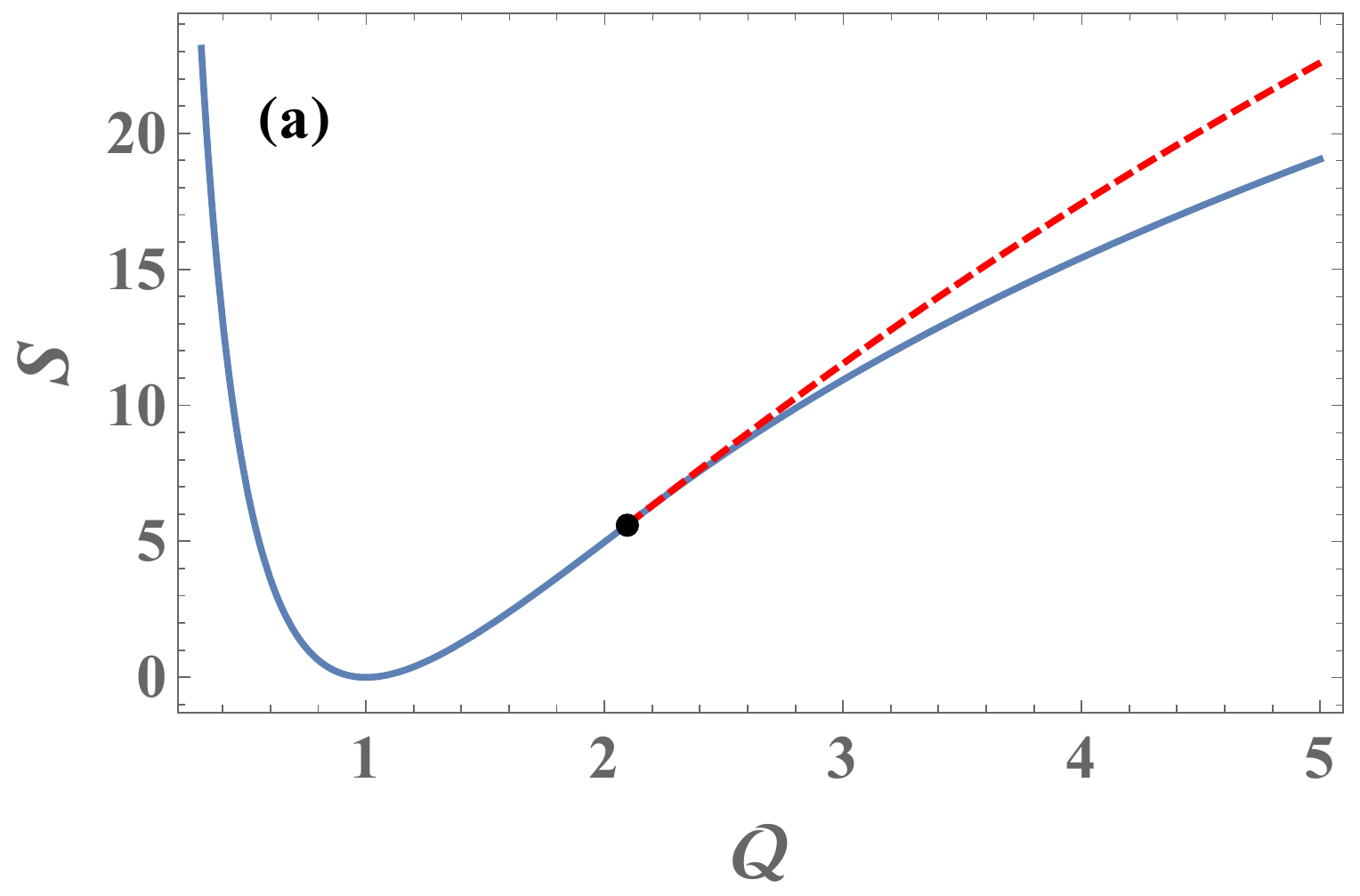}
\includegraphics[width=0.492\linewidth,clip=]{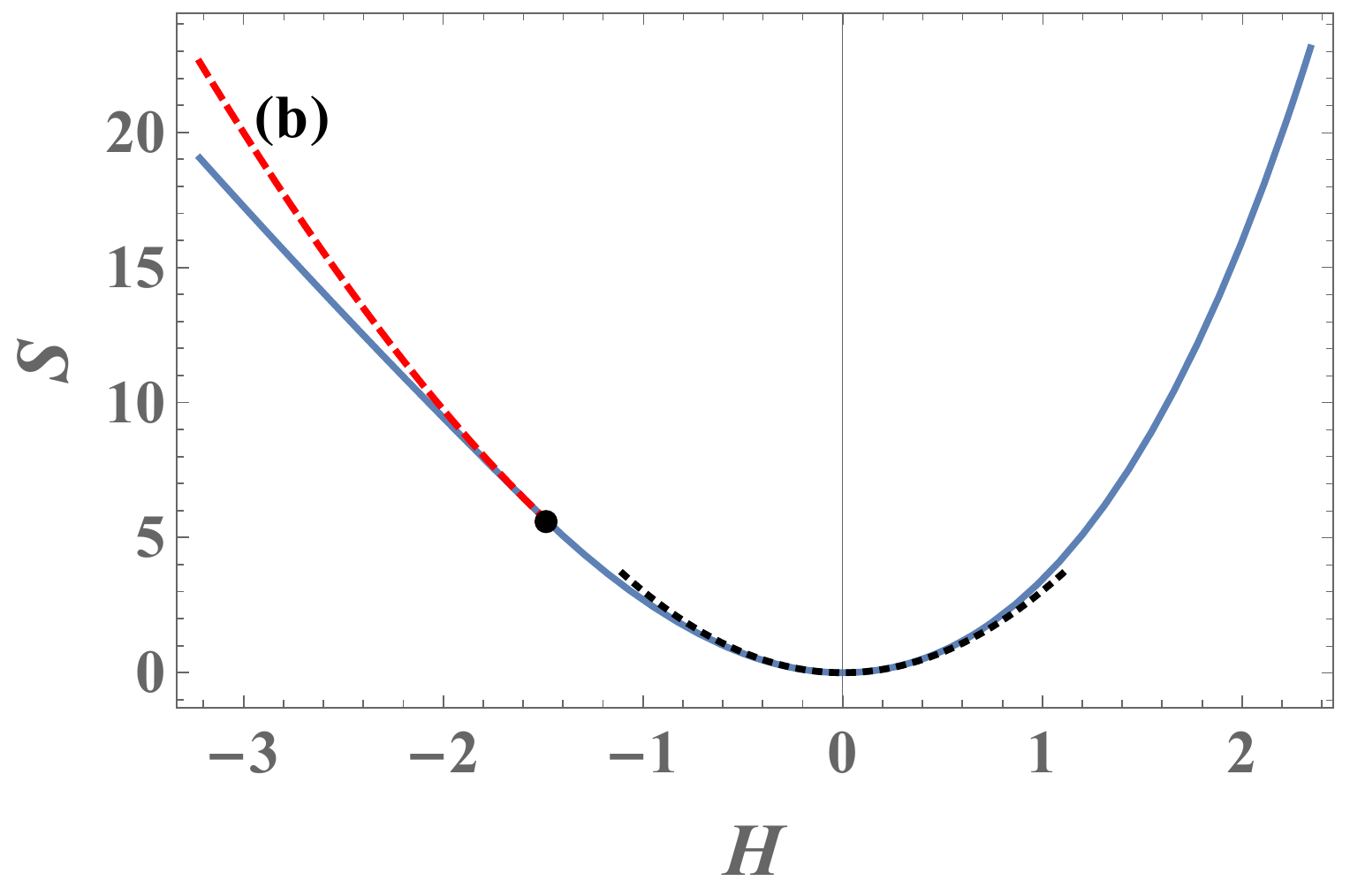}
\caption{$S$ vs. $\mathcal{Q}$ (a) and vs. $H$ (b). There is a second-order dynamical phase transition at the critical value $\mathcal{Q}_c = 2.165\dots$, corresponding to $H_c = -2 \ln \mathcal{Q}_c = -1.544\dots$.
The dashed line represents the continuation of the non-optimal branch of $S$, which corresponds to mirror-symmetric interface histories $h(x,t)$. The critical point $\mathcal{Q}_c$ is also the inflection point of the non-optimal branch of $S(\mathcal{Q})$, i.e., the point at which $d^{2}S/d\mathcal{Q}^{2}$ changes sign.
The dotted line in (b) corresponds to the parabolic approximation of $S(H)$ around $H=0$, given by the first line of Eq.~\eqref{SofHasymptotic}.}
\label{fig:SofQandofH}
\end{figure}

By now it should be clear that $F$ in fact plays the role of a Landau free energy. Thus, the DPT of $S(\mathcal{Q})$ is of second order, i.e., the second derivative $d^{2}S/d\mathcal{Q}^{2}$ jumps at the critical point.
$S$ is plotted as a function of $\mathcal{Q}$ and as a function of $H$ in Figs.~\ref{fig:SofQandofH} (a) and (b) respectively.

In Appendix \ref{app:asymptotic}, we show that the asymptotic behaviors of $S(H)$ are
\be
\label{SofHasymptotic}
S\left(H\right)\simeq\begin{cases}
\frac{\sqrt{\pi}\,H^{2}}{2-\sqrt{2}}, & \left|H\right|\ll1,\\[2mm]
18e^{H/2}, & H\gg1,\\[2mm]
\frac{8\sqrt{2}}{3\sqrt{3}}\left(-H\right)^{3/2}, & -H\gg1.
\end{cases}
\ee
The first and third lines follow straightforwardly from an application of the formula \eqref{SofHFromMS} to analogous results of \cite{MeersonSchmidt17}, but the second line is nontrivial and displays a very different scaling behavior to the power law $S \sim H^{5/2}$ that has been observed in this tail for the noisy ($D>0$) KPZ equation.
This difference is not so surprising, when recalling that in \cite{MeersonSchmidt17} it was found that, at $D > 0$, the $H \gg 1$ tail is dominated by the noise in the dynamics, and the fluctuations of the initial condition do not contribute in the leading order.

\begin{figure}[ht]
\includegraphics[width=0.65\linewidth,clip=]{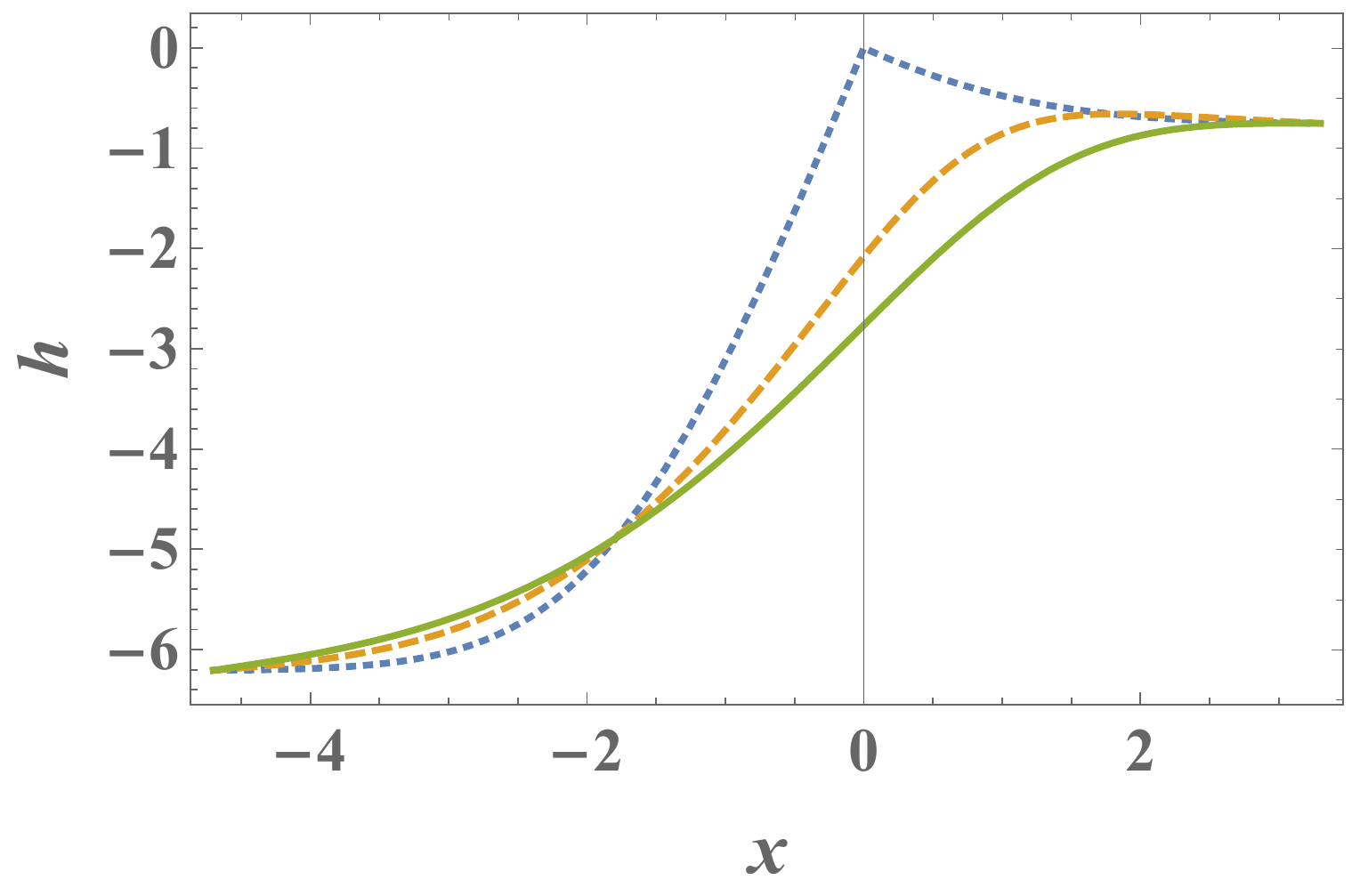}
\caption{The time-dependent optimal history $h(x,t)$ at (rescaled) times $t=0, 1/2, 1$ (dotted, dashed and solid lines, respectively), conditioned on reaching height $H = -2.76$. 
Since $H < H_c$ is supercritical, the optimal history spontaneously breaks the mirror symmetry $x \leftrightarrow -x$. There is an additional optimal path with the exact same action which is given by the mirror image of the displayed solution.}
\label{fig:hofxAsym}
\end{figure}

The optimal path $h(x,t)$ conditioned on a given $H$ is found exactly, as follows. One first calculates $h_0(x)$ separately for the left ($x<0$) and right ($x>0$) halves of the system, as described above (at $\mathcal{Q} > \mathcal{Q}_{c}$).
The (symmetry-broken) optimal path for $H = -2.76$ is plotted in Fig.~\ref{fig:hofxAsym}.

\medskip

\medskip

\section{Discussion}

\label{sec:disc}

We have analytically calculated the exact large-deviation function $S(H)$ that describes the distribution of the single-point height $H$ for a relaxing ($D=0$) KPZ interface in 1+1 dimension at short times, starting from a Brownian initial condition. The most remarkable feature of $S(H)$ is its second-order dynamical phase transition, which we describe by formulating an effective Landau theory, with a Landau free energy that we too calculate exactly.

We found that the mechanism that leads to the occurrence of the dynamical phase transition in our problem is as follows. Using the parametrization $\mathcal{Q} = e^{-H/2}$, we find that $\mathcal{Q}$ can be written as the sum of two i.i.d. random variables that correspond to the contributions of the left- and right-halves of the initial interface. We then found the large-deviation function $s(\mathcal{Q}_R)$ that describes the distribution of each of these random variables and showed that is non-convex. It is this non-convexity that ultimately leads to the breaking of the mirror symmetry $x \leftrightarrow -x$ of the interface at supercritical $H$, and in fact, we find that the critical point of $S(H)$ is related to the inflection point of $s(\mathcal{Q}_R)$.

In addition, the simplification that is obtained by using the Hopf-Cole solution to the KPZ relaxation dynamics enabled us to calculate the exact optimal history of the system $h(x,t)$ conditioned on reaching a given height $H$. In principle, this has also been achieved for the (noisy) KPZ equation in several other settings recently by using the inverse scattering method \cite{KLD21, KLD22}. However, our results are more explicit and do not require more technically involved steps such as inverting the scattering transform and/or using numerical algorithms to evaluate a Fredholm operator inversion formula.

From our results it is easy to obtain the short-time relaxation height statistics for a KPZ interface in other settings that are similar to the one we considered here:
\begin{enumerate}[wide, labelwidth=!, labelindent=0pt]
\item
For a relaxing KPZ interface in half space, $x\ge 0$, with a Neumann boundary condition at the origin, $\partial_{x}h\left(x=0,t\right)=0$, and with an initial condition that is given by a (one-sided) Brownian motion, the OFM problem yields a problem that can be solved by using the method of mirrors. This solution corresponds to the symmetric (analytic) branch of the large-deviation function that is given in Eq.~\eqref{SofQtwices}. However, the action is smaller by a factor of 2 than in the full-space problem because it is evaluated only on the half space. Therefore, the large-deviation function would simply be $S_{\text{hs}}\left(\mathcal{Q}\right)=s\left(\mathcal{Q}/2\right)$.
\item
For a relaxing KPZ interface in full space, $-\infty < x < \infty$, whose initial condition has a flat left half ($x<0$) and a right half is given by a one-sided Brownian motion, one simply finds that $\mathcal{Q}=\mathcal{Q}_{L}+\mathcal{Q}_{R}=1/2+\mathcal{Q}_{R}$. The corresponding large-deviation function is therefore simply
$S_{\text{flat-BM}}\left(\mathcal{Q}\right)=s\left(\mathcal{Q}-1/2\right)$.
Similarly, for an initial condition that is a droplet at $x<0$, $h(x<0) = -Ax$ with $A \to \infty$, and Brownian at $x>0$ one has $\mathcal{Q}_{L} = 0 $ so $\mathcal{Q}=\mathcal{Q}_{R}$, and the corresponding large-deviation function is simply given by $s(\mathcal{Q})$.
\end{enumerate}

One could try to extend our results to other types of random initial conditions, such as fractal Brownian motion, for which the OFM was recently formulated in \cite{MO22}.
Finally, it would be interesting to study the long-time relaxation statistics, in the regimes of typical fluctuations and large deviations.
One would expect, in analogy with the known results for the noisy KPZ equation \cite{LDlongtime, SMP, KLD18, CorwinGhosal, CGKLDT18} (see also \cite{DL98, DA99, BD00, MP18}), that sufficiently far into the distribution tails, our predictions would still hold, at arbitrary times.
Another interesting outstanding open question, for $D>0$, is that of finding the critical curve $H_c$ as a function of $\sigma$. We now know three points on this curve: at $\sigma =0, 1$ and $\infty$.

\bigskip

\section*{ACKNOWLEDGMENTS}

I thank Baruch Meerson for useful discussions and for a critical reading of the paper.

\bigskip

\appendix

\section{Asymptotic limits}
\label{app:asymptotic}
\renewcommand{\theequation}{A\arabic{equation}}
\setcounter{equation}{0}

In this Appendix we obtain the asymptotic behaviors of $S(H)$ that are given in Eq.~\eqref{SofHasymptotic} of the main text. As stated there, the behaviors at $|H| \ll 1$ and at $H \to -\infty$ follow immediately from more general results of \cite{MeersonSchmidt17}, for which we provide some more details now, whereas the $H \to \infty$ behavior is derived below and is very different to behaviors that were previously observed in similar settings.
The reason for this difference is briefly explained in the main text, shortly after Eq.~\eqref{SofHasymptotic}.

To remind the reader, our large-deviation function $S(H)$ is related to the function $\mathcal{S}\left(H,\sigma\right)$ from \cite{MeersonSchmidt17} via Eq.~\eqref{SofHFromMS}.
$\mathcal{S}\left(H,\sigma\right)$ is not known exactly for general $\sigma>0$, but its asymptotic behaviors were found in \cite{MeersonSchmidt17} to be
\be
\label{SofHsigmaAsymptotic}
\mathcal{S}\left(H,\sigma\right)\simeq\begin{cases}
\frac{\sqrt{\pi}\,H^{2}}{\sqrt{2}+\left(2-\sqrt{2}\right)\sigma^{2}}, & \left|H\right|\ll1,\\[3mm]
\frac{4\sqrt{2}}{15\pi}H^{5/2}, & H\gg1,\\[3mm]
\frac{8\sqrt{2}}{3\sqrt{3\sigma^{4}+1}}\left(-H\right)^{3/2}, & -H\gg1 .
\end{cases}
\ee
The first and third lines in Eq.~\eqref{SofHasymptotic} of the main text follow immediately by using \eqref{SofHFromMS} in the first and third lines of \eqref{SofHsigmaAsymptotic}. However, one cannot do this for the second line because, as it turns out, the limits $H \to \infty$ and $\sigma \to \infty$ do not commute.
Indeed, the second line of \eqref{SofHsigmaAsymptotic} was obtained in \cite{MeersonSchmidt17} by neglecting the contribution of fluctuations of the initial condition. It was found that, at any $D > 0$, this tail is dominated by the noisy dynamics, while the fluctuations of the initial condition do not contribute. This result cannot be valid in the case $D=0$ that we consider here.
We must therefore use a more careful approach.

\begin{figure}[ht]
\includegraphics[width=0.3\textwidth,clip=]{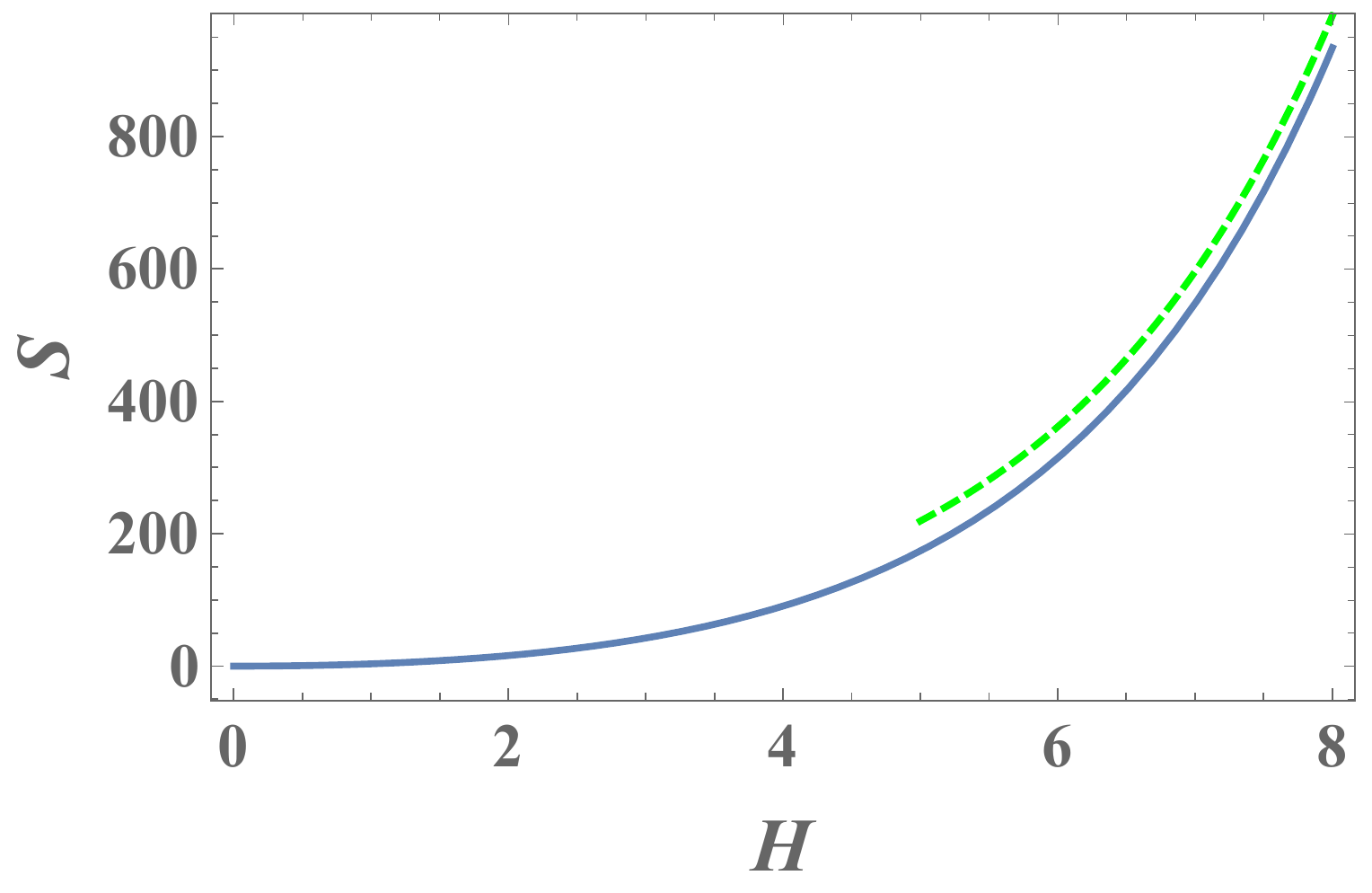}
\caption{The $H \to \infty$ tail of the large-deviation function $S(H)$, together with its approximation \eqref{SofHHtoinfty} (dashed line).}
\label{fig:SofHtail}
\end{figure}

We now derive the $H \to \infty$ behavior of $S(H)$, as given in the second line of Eq.~\eqref{SofHasymptotic} of the main text.
This limit corresponds to $\Lambda \to -\infty$. 
In order to obtain the asymptotic behavior of $S(H)$ in this limit, it is useful to analyze the effective potential $U(\psi)$ from Eq.~\eqref{Upsidef}. Note that, at any $\Lambda < 0$, the (global) maximum of $U(\psi)$ is at $\psi_1=2\ln\left(-\frac{\Lambda}{4\sqrt{4\pi}}\right)$, where the value
$U_{\max} = U(\psi_1) = -2\ln\left(-\frac{\Lambda}{4\sqrt{4\pi}}\right)-2$ is reached.
At sufficiently large $|\Lambda|$, one has $\psi_1 > 0$ and therefore, to be consistent with the boundary conditions $\psi(0)=0$ and $\psi(\infty)=\infty$, the `energy' $E$ of the effective particle must be larger than $U_{\max}$.
In addition, it turns out that $E$ is negative for $\Lambda < 0$.
On the other hand, at $-\Lambda \gg 1$, one has $-U_{\max} \ll -\Lambda$, which therefore implies
$0 < -E  \ll -\Lambda$.
Comparing this with the expression \eqref{EofQR}, we therefore must require that the two terms on the right-hand side of \eqref{EofQR} cancel each other out in the leading order, i.e., 
\be
\frac{\Lambda^{2}\mathcal{Q}_{R}^{2}}{32}\simeq-\frac{\Lambda}{2\sqrt{4\pi}}
\ee
which yields
\be
\Lambda\left(\mathcal{Q}_{R}\ll1\right)\simeq-\frac{8}{\sqrt{\pi}\,\mathcal{Q}_{R}^{2}}
\ee
using $ds/d\mathcal{Q}_R = \Lambda$, we integrate the last equation with respect to $\mathcal{Q}_R$ and obtain
\be
s\left(\mathcal{Q}_{R}\ll1\right)\simeq\frac{8}{\sqrt{\pi}\,\mathcal{Q}_{R}} \, .
\ee
Now, using $S\left(\mathcal{Q}\right)=2s\left(\mathcal{Q}_{R}=\frac{\mathcal{Q}}{2}\right)$ (which is valid at all $\mathcal{Q}<\mathcal{Q}_{c}$), we obtain
\be
S\left(\mathcal{Q}\to0\right)\simeq\frac{32}{\sqrt{\pi}\,\mathcal{Q}}
\ee
or, in terms of $H$
\be
\label{SofHHtoinfty}
S\left(H\to\infty\right)\simeq18e^{H/2}.
\ee
The asymptotic formula \eqref{SofHHtoinfty} is compared to the exact function $S(H)$  in Fig.~\ref{fig:SofHtail}, showing good agreement at large $H$.
Eq.~\eqref{SofHHtoinfty} grows with $H$ much faster than in the $\sim H^{5/2}$ tail that has been observed in the (noisy) KPZ equation in many other contexts. 



\begin{thebibliography}{99}
\bibitem{signlambda}
Changing the sign of $\lambda$ is  equivalent to changing the sign of $h$.
\end{thebibliography}
\end{document}